\documentclass[sigconf]{acmart}

\usepackage{caption}
\usepackage{subcaption}

\usepackage[normalem]{ulem}
\useunder{\uline}{\ul}{}

\newcommand{\chatgpt}{{\tt ChatGPT}}

\AtBeginDocument{%
  }

\setcopyright{acmlicensed}
\copyrightyear{2018}
\acmYear{2018}
\acmDOI{XXXXXXX.XXXXXXX}

\acmConference[Conference acronym 'XX]{Make sure to enter the correct
  conference title from your rights confirmation emai}{June 03--05,
  2018}{Woodstock, NY}
\acmISBN{978-1-4503-XXXX-X/18/06}

\begin{document}

\title{Can \chatgpt\ Pass a Theory of Computing Course?}

\author{Matei A. Golesteanu}
\authornote{The first two authors contributed equally to this research.}
\email{matei.golesteanu@westpoint.edu}
\affiliation{%
  \institution{United States Military Academy}
  \city{West Point}
  \state{New York}
  \country{USA}
}

\author{Garrett B. Vowinkel}
\email{garrett.vowinkel@westpoint.edu}
\affiliation{%
  \institution{United States Military Academy}
  \city{West Point}
  \state{New York}
  \country{USA}
}

\author{Ryan E. Dougherty}
\email{ryan.dougherty@westpoint.edu}
\orcid{0000-0003-1739-1127}
\affiliation{%
  \institution{United States Military Academy}
  \streetaddress{601 Thayer Road}
  \city{West Point}
  \state{New York}
  \country{USA}
  \postcode{10996}
}

\renewcommand{\shortauthors}{Golesteanu, Vowinkel, and Dougherty}

\begin{abstract}
Large Language Models (LLMs) have had considerable difficulty when prompted with mathematical questions, especially those within theory of computing (ToC) courses. 
In this paper, we detail two experiments regarding our own ToC course and the \chatgpt\ LLM.
For the first, we evaluated \chatgpt's ability to pass our own ToC course's exams.
For the second, we created a database of sample ToC questions and responses to accommodate other ToC offerings' choices for topics and structure. 
We scored each of \chatgpt's outputs on these questions. 
Overall, we determined that \chatgpt can pass our ToC course, and is adequate at understanding common formal definitions and answering ``simple''-style questions, e.g., true/false and multiple choice. However, \chatgpt\ often makes nonsensical claims in open-ended responses, such as proofs.
\end{abstract}

\ccsdesc[500]{Theory of computation~Formal languages and automata theory}
\ccsdesc[500]{Social and professional topics~Computing education}
\ccsdesc[300]{Mathematics of computing~Discrete mathematics}

\keywords{theoretical computer science, computer science education, large language model, formal languages, automata theory, chatgpt}

\received{20 February 2007}
\received[revised]{12 March 2009}
\received[accepted]{5 June 2009}

\maketitle

\section{Introduction}

Large Language Models (LLMs) have had a serious impact on student academic integrity, including Computer Science. 
Many LLMs train based upon publicly available data, which is proportional to the popularity of students taking courses in certain topics.
In terms of CS, the most popular topic is introductory programming.
However, for less popular topics, do LLMs perform as well, especially in technical topics that require precise arguments and are rigorous?

This paper attempts to answer this question for a Theory of Computing (ToC) course, which is a required course for nearly all undergraduate CS curricula. 
It is a traditionally difficult course as it nearly always requires students' be able to write precise proofs and create mathematical constructions.
We tested the \chatgpt\ LLM\footnote{\url{http://chatgpt.com/}} within the ToC course at our institution in two different experiments. 
First, we tested its ability to pass our course's exams by grading its responses as though it were an actual student.
Second, we created a large database of representative questions other offerings of ToC may ask, and determine \chatgpt's ability to answer those questions.

Our methods attempt to model what a student would do if they were in a ToC course and used \chatgpt.
The goals of this work are to (1) inform educators on \chatgpt's ability to pass a ToC course, (2) also inform eductors what they can do to limit students' ability to use \chatgpt\ without their learning ToC concepts, and (3) if desired, determine where \chatgpt\ struggles to help inform future ToC course designs on how to make LLM tools be more effective in the classroom.
The remainder of this paper is organized as follows.
\begin{itemize}
    \item Section~\ref{sec:background} contains related work to ours. 
    \item Section~\ref{sec:method} contains our method, which involves these two experiments.
    \item Section~\ref{sec:results} contains our results to our two experiments.
    \item Section~\ref{sec:discussion} discusses our results that will inform ToC educators of what aspects of ToC \chatgpt\ struggled with the most.
    \item Section~\ref{sec:threats} contains our threats to validity.
    \item Section~\ref{sec:future_work} contains our recommendations to educators and future work.
    \item Section~\ref{sec:conclusion} concludes the paper. 
\end{itemize}


\section{Background Research}\label{sec:background}
As far as we are aware, there has not been any prior research with \chatgpt's ability to pass a ToC course.
However, there has been several works regarding other common CS courses. 
Sharpe et al. \cite{sharpe2024can} recently determined that \chatgpt\ can easily pass a CS1 course.
Banerjee et al. \cite{Banerjee07} discussed ``different opportunities with use cases of \chatgpt\ in augmenting the learning experience.'' 
The paper proposes several methods in order to integrate the use of AI’s into the classroom, such as having students explain an answer given by \chatgpt, or explain the logic used by \chatgpt. 
Echeverría et al. \cite{10478897} conducted an analysis of \chatgpt's performance in computer engineering exams, and found that \chatgpt\ can effectively pass that class even though it contained many different types of questions.

Although not directly a CS topic, a closely related work to ours is that of Frieder et al. \cite{frieder2024mathematical}, about \chatgpt's ability to answer discrete math questions; this course is very often a pre-requisite for ToC.
They determined that \chatgpt\ can reasonably well solve undergraduate-level problems, but not graduate-level ones. 
Relevant to our work, they used \chatgpt\ version 4, as it is more accurate than previous versions of \chatgpt; we will be using this version.



Research was also done into using ``prompt engineering'' in order to tailor \chatgpt's responses to make them more accurate. 
Vel\'asquez-Henao et al. \cite{Henao23} describe ``a methodology for optimizing interactions with Artificial Intelligence language models''. 
They found that by using specific, creatively framed questions for \chatgpt, they could effectively ``elicit precise, accurate, and contextually appropriate responses.''
We adopt a similar approach of prompt engineering to improve \chatgpt's responses. 

\section{Course Context and Method}\label{sec:method}

Our institution is a small liberal-arts college (SLAC) located in the northeastern United States, with between 30 and 50 CS students graduating every year. 
Our ToC course consists of five main sections: (1) regular languages, (2) context-free languages, (3) Turing Machines and Decidability, (4) Undecidability, and (5) NP-Completeness.
The formal pre-requisites are discrete mathematics and data structures, a CS2 course; an informal pre-requisite is programming languages as it helps with some formal notation.
There are no formal post-requisites, but students very often take a year-long capstone course and an Operating Systems course in the next semester.

For both experiments and the results we give below, there are many acronyms and phrases that we use for terms in our ToC course, which are very standard.
For simplicity, Table~\ref{tbl:acronyms} has a mapping from acronym or shortened phrase to its meaning; for a refresher on the formal definitions of these terms, see Sipser \cite{sipser2013introduction}.

\begin{table}[]
\caption{\label{tbl:acronyms}Acronyms and shortened phrases used throughout our ToC course and this paper.}
\begin{tabular}{l|l}
\textbf{Acronym} & \textbf{Meaning}                  \\ \hline
CFG              & Context-Free Grammar              \\
Count            & Countability                      \\
CNF              & Chomsky Normal Form               \\
Dec.             & Decidable                         \\
DFA              & Deterministic Finite Automaton    \\
DPDA             & Deterministic Pushdown Automaton  \\
Induct           & Structural Induction              \\
NFA              & Nondeterministic Finite Automaton \\
NP               & Nondeterministic Polynomial Time  \\
PDA              & Pushdown Automaton                \\
PL               & Pumping Lemma                     \\
Rec.             & Recognizable                      \\
Regex            & Regular Expression                \\
Rice             & Rice's Theorem                    \\
TM               & Turing Machine                    \\
Undec.           & Undecidable                       
\end{tabular}
\end{table}

\subsection{Experiment 1: \chatgpt\ in our ToC Course}
Student assessment for our course was divided up as follows:
\begin{itemize}
    \item Paper Writing: 450 points
    \item Two Midterm Exams: 200 points total
    \item Formal Group Presentation: 100 points
    \item Final Exam: 250 points
\end{itemize}

The paper writing assessment is an extension based on the work of Dougherty \cite{dougherty2024experiences}, involving writing a research paper in small groups across the semester using knowledge gained in our ToC course.
The formal group presentation is based on the fifth section of our ToC course, namely NP-Completeness.
Students had to give a full proof based on a given proof from the Garey and Johnson textbook \cite{garey1979computers}, which omits some details.
Finally, the midterms and final are standard in-class exams.

In Dougherty's method (and ours), the paper topics are different for each group; the same is true of the formal group presentation.
Therefore, we determine whether \chatgpt\ can adequately pass the three exams.
The results we find would be applicable to many other institutions and course contexts because many ToC courses have a very heavy weight to exams; further, these exams test students' baseline understanding of the material. 

The first midterm exam only covered Section 1 (regular languages), the second exam only covered Section 2 (context-free languages), and the final exam covered all 5 sections.
As for individual topics, Midterm 1 contained questions about induction, DFAs, NFAs, NFA to DFA conversion, the Pumping Lemma for Regular Languages, and regular expressions.
Midterm 2 contained questions about Context-Free Grammars, Pushdown Automata, conversions between these two models, the Pumping Lemma for Context-Free Languages, and Deterministic Pushdown Automata.
And finally, the Final Exam contained questions including topics from Midterms 1 and 2, and Turing Machines, Countability, Decidability, Undecidability, and NP-Completeness. 

\subsection{Experiment 2: Question Database}

We do note that ToC course contexts do vary in terms of topics tested, even within the five sections mentioned above.
For example, some ToC courses may teach the Pumping Lemma for Regular Languages, and some may instead choose the Myhill-Nerode equivalence relation. 
Further, the three exams may not be as comprehensive for detailing specifically in which concepts \chatgpt\ performs well as they are a limited-time environment.
Therefore, in addition to the results we find for our own specific ToC course, we also compiled many different ToC questions to cover a larger spectrum of what other ToC offerings may have.

We focus on the first four sections of our ToC as these sections on a macro level are very widespread across ToC course offerings.
Here, we created 450 questions about basic topics in ToC, evenly divided among (1), (2), and (3) + (4).
These questions were evenly divided amongst True/False, Multiple Choice, and Proof question types.
The design of these questions is suitable for an undergraduate-level ToC course; some are based on definitions of formal objects, some on simple properties of the given language or model, asking \chatgpt\ to provide examples of a language/model, etc.
Here is a sample of our generated questions:
\begin{itemize}
    \item "Is the complement of a regular language also a regular language? Justify your answer."
    \item "True or false: No arbitrarily long language can ever be regular."
    \item "Prove that the language $L2 = \{0^a1^b : a > b\}$ is not regular using the pumping lemma."
    \item "What is the formal definition of a Deterministic Finite Automata?"
    \item "Give an example of a language that a DFA cannot model?"
    \item "True or false: There exists an NFA such that it cannot be converted to a DFA."
    \item "Give a recursive definition for the following set: the set $X$ of all strings in $\{0, 1\}^\star$ containing 00."
    \item "True or False: the transition function of a pushdown automata is the same as that of an NFA."
    \item "True or false: the pumping lemma for context free languages can be used to shown that a language is context free.''
    \item "What is the formal definition of a context free grammar?''
\end{itemize}

We then fed all of the questions into \chatgpt\ using a set precursor message before every prompt: 
``I will ask you a question to test your academic theoretical computer science knowledge, answer to the best of you abilities. Here it is:''. 
After receiving and recording the responses, we assigned them a grade from 0--4 based on a rubric. 
It is important to note that we did not grade simply the response in the case of True/False or Multiple Choice questions, but also the reasoning and correctness behind the responses. 
Therefore, it is possible that it chose the correct option, but did not earn a 4/4 due to the reasoning/correctness. 
The rubric is as follows:
\begin{itemize}
  \item 4: Totally correct and complete solution, covering all aspects of the question, and answered with enough depth to warrant full points on an undergraduate-level ToC exam.
  \item 3: Mostly correct and mostly complete solution. This covers answers that are mostly correct, but have minor mistakes; these can include not fully answering minor parts of the question, slight missteps in a proof, or otherwise making some small error which does not detract from the overall answer, but would also not earn full points on an undergraduate-level ToC exam.
  \item 2: This grade will be given to responses that have a somewhat correct answer but include a major flaw. This could include proofs that arrive to a correct conclusion or mostly adhere to a correct logical flow, but make some large incorrect assumption or step which detracts from the overall value. 
  \item 1: This grade is the equivalent of a 3 but on the opposite side, being a mostly incorrect or incomplete solution. The grade covers responses which have some semblance of truth and correctness, but still make a large error in either proof or justification.
  \item 0: Completely wrong answer. This grade will be given to a response if it is absolutely incorrect and does not even begin to contain some truthful aspect or a path towards the correct answer. 
  
\end{itemize}

\chatgpt\ was then given the opportunity to try again if it did not score a 4/4. 
Basic feedback given to \chatgpt\ in the following form: ``You got a x/4, improve the answer.'', with x being the appropriate rubric score. 
A maximum of three attempts were given for each question. 
The feedback is purposefully consistent and vague.

\subsection{Experimental Setup}

In both experiments, none of the questions used deception to induce \chatgpt\ to produce false answers.
We used \chatgpt\ version 4 as it is the most accurate currently available model and we could upload a document directly.
Many of the questions require mathematical notation; see the sample questions above for examples.
However, an easily readable notation is not directly compatible with ASCII characters normally typed in on a keyboard.
For Experiment 1, as we are using \chatgpt\ 4, we directly uploaded the exams as Word documents to \chatgpt\ to avoid this issue; for realism purposes, these exams were identical to those given to the ToC students at our institution.
For Experiment 2, we formatted the questions as close as possible to the formal notation with ASCII characters; no question in this data set required a diagram for the prompt.
For example, consider the third example question above about showing that $L2 = \{0^a 1^b : a > b\}$ is not regular; what we instead used for this prompt was \verb|L2 = {0^a 1^b : a > b}|.
In both experiments, only a few times did \chatgpt\ not understand the initial prompt, and every issue was corrected with a clarifying prompt.
One example is in Experiment 1 when \chatgpt\ interpreted $\sigma^i$ for $\sigma_i$.

\section{Results}\label{sec:results}

Table~\ref{tbl:experiment1_result_summary} contains our overall results for Experiment 1. 
For each of the three exams, it contains the scores for both the verbatim questions in the exam and revised answers after trying again when given a hint as to why the answer it gave is wrong.
Overall, \chatgpt\ performed well on all three exams, earning over 80\% in all experiments, and up to 93\% after re-tries with helpful hints.
The weighted average was equivalent at our institution to a {\tt B-} as the initial letter grade, and an {\tt A-} after re-tries.
Each of the exams contains ``main'' questions and ``bonus'' questions; the listed scores in our results include these bonus questions for accuracy.

\begin{table}
\caption{\label{tbl:experiment1_result_summary}Overall results for Experiment 1.}
\begin{tabular}{l|l|l|}
\cline{2-3}
                                                 & Initial & After Retry \\ \hline
\multicolumn{1}{|l|}{Midterm 1 (Section 1)}      & 80.5\%        & 93.0\%             \\ \hline
\multicolumn{1}{|l|}{Midterm 2 (Section 2)}      & 87.5\%        & 92.0\%             \\ \hline
\multicolumn{1}{|l|}{Final Exam (Sections 1--5)} & 81.0\%        & 88.6\%             \\ \hline
\multicolumn{1}{|l|}{\textbf{Average}}           & 82.33\%       & 90.33\%            \\ \hline
\end{tabular}
\end{table}

Table~\ref{tab:results} contains our overall results for Experiment 2.
It shows the average best score for each topic tested. 
Here, ``Average Max'' refers to the mean of all of the scores, where the score for each question is the best attempt on that particular question.
Overall, \chatgpt\ earned an 85\% for Experiment 2; this is 91.5\% on true/false questions, 87.3\% on multiple choice questions, and 78.8\% on proof-type questions.

\begin{table*}
\caption{\label{tab:results}Results for Experiment 2; we show the average maximum score obtained for each of the 450 questions we asked \chatgpt\ including re-tries, as well as the average across question types and categories.
Indicated cells show when the average grade is below failing at our institution.
For the meaning of each of the acronyms and short phrases, see Table~\ref{tbl:acronyms}.
}
\begin{tabular}{l|lllll|lllll|lll|ll|ll}
\cline{2-16}
                                    & \multicolumn{5}{c|}{Section 1}                                                                                                           & \multicolumn{5}{c|}{Section 2}                                                                                                                                           & \multicolumn{3}{c|}{Section 3}                                                                            & \multicolumn{2}{c|}{Section 4}                     &                                    &  \\ \cline{2-17}
                                    & \multicolumn{1}{c|}{Induct} & \multicolumn{1}{c|}{DFA}  & \multicolumn{1}{c|}{NFA}  & \multicolumn{1}{c|}{PL}   & \multicolumn{1}{c|}{Regex}  & \multicolumn{1}{c|}{CFG}                         & \multicolumn{1}{c|}{PDA}                         & \multicolumn{1}{c|}{CNF}  & \multicolumn{1}{c|}{DPDA} & \multicolumn{1}{c|}{PL} & \multicolumn{1}{c|}{TMs}                         & \multicolumn{1}{c|}{Count} & \multicolumn{1}{c|}{Dec., Rec.} & \multicolumn{1}{c|}{Undec.} & \multicolumn{1}{c|}{Rice} & \multicolumn{1}{c|}{\textbf{Avg.}} &  \\ \cline{1-17}
\multicolumn{1}{|l|}{T/F}           & \multicolumn{1}{l|}{3.7}       & \multicolumn{1}{l|}{4.0}  & \multicolumn{1}{l|}{3.4}  & \multicolumn{1}{l|}{3.8}  & 3.4                         & \multicolumn{1}{l|}{4.0}                         & \multicolumn{1}{l|}{3.8}                         & \multicolumn{1}{l|}{3.6}  & \multicolumn{1}{l|}{3.7}  & 3.6                     & \multicolumn{1}{l|}{3.6}                         & \multicolumn{1}{l|}{3.6}   & 3.4                             & \multicolumn{1}{l|}{3.7}    & 3.6                       & \multicolumn{1}{l|}{3.66}          &  \\ \cline{1-17}
\multicolumn{1}{|l|}{MC}            & \multicolumn{1}{l|}{3.8}       & \multicolumn{1}{l|}{4.0}  & \multicolumn{1}{l|}{3.3}  & \multicolumn{1}{l|}{3.1}  & 3.4                         & \multicolumn{1}{l|}{3.2}                         & \multicolumn{1}{l|}{3.9}                         & \multicolumn{1}{l|}{3.5}  & \multicolumn{1}{l|}{3.5}  & 3.7                     & \multicolumn{1}{l|}{3.3}                         & \multicolumn{1}{l|}{3.6}   & 3.0                             & \multicolumn{1}{l|}{3.3}    & 3.7                       & \multicolumn{1}{l|}{3.49}          &  \\ \cline{1-17}
\multicolumn{1}{|l|}{Proof}         & \multicolumn{1}{l|}{3.2}       & \multicolumn{1}{l|}{3.7}  & \multicolumn{1}{l|}{3.0}  & \multicolumn{1}{l|}{3.4}  & {\ul \textit{\textbf{2.4}}} & \multicolumn{1}{l|}{{\ul \textit{\textbf{2.5}}}} & \multicolumn{1}{l|}{{\ul \textit{\textbf{2.5}}}} & \multicolumn{1}{l|}{3.6}  & \multicolumn{1}{l|}{3.4}  & 3.4                     & \multicolumn{1}{l|}{{\ul \textit{\textbf{2.4}}}} & \multicolumn{1}{l|}{3.5}   & 3.4                             & \multicolumn{1}{l|}{3.4}    & 3.4                       & \multicolumn{1}{l|}{3.15}          &  \\ \cline{1-17}
\multicolumn{1}{|l|}{\textbf{Avg.}} & \multicolumn{1}{l|}{3.57}      & \multicolumn{1}{l|}{3.90} & \multicolumn{1}{l|}{3.23} & \multicolumn{1}{l|}{3.43} & 3.07                        & \multicolumn{1}{l|}{3.23}                        & \multicolumn{1}{l|}{3.40}                        & \multicolumn{1}{l|}{3.57} & \multicolumn{1}{l|}{3.53} & 3.57                    & \multicolumn{1}{l|}{3.10}                        & \multicolumn{1}{l|}{3.57}  & 3.27                            & \multicolumn{1}{l|}{3.47}   & 3.57                      &                                    &  \\ \cline{1-16}
\end{tabular}
\end{table*}


\section{Discussion}\label{sec:discussion}


\subsection{Experiment 1 Discussion}

For brevity we will not include a discussion of each question individually, but rather include any discussion that will be of use for educators and/or gave surprising results.



\begin{figure*}
    \centering
    \begin{subfigure}{.35\textwidth}
      \centering
      \includegraphics[width=1.0\textwidth]{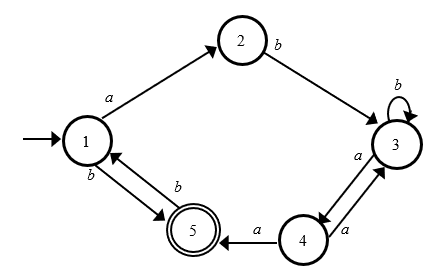}
      \caption{Original NFA for Midterm 1's third question.}
      \label{fig:original_nfa}
    \end{subfigure}
    \qquad\qquad
    \begin{subfigure}{.50\textwidth}
      \centering
      \includegraphics[width=1.0\textwidth]{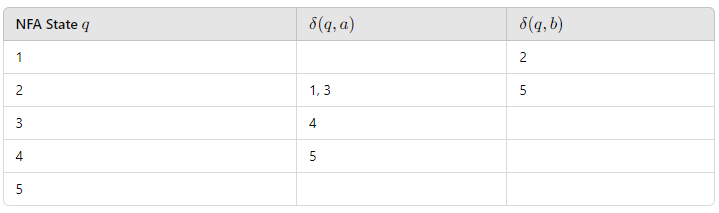}
      \caption{Initial Table from \chatgpt.}
      \label{fig:initial_table_for_nfa}
    \end{subfigure}
    \caption{Example of \chatgpt's issue with reading state diagrams directly made in Word.}
    \label{fig:midterm1_q3}
\end{figure*}

\chatgpt\ could read state diagram pictures correctly, but this heavily depended on the picture's format.
For example, in the induction question for Midterm 1, students needed to observe a state diagram of a DFA given as a JPG picture included in the exam Word document; here, \chatgpt\ read that picture correctly.
However, in the same exam's third question, about a standard NFA to DFA conversion, the state diagram was directly drawn in Word using its built-in shapes, and not as a JPG image.
As a result, \chatgpt\ was unable to correctly read the picture even after a re-try, although some of the transitions and state names were correct; see Figure~\ref{fig:midterm1_q3} for the initial table.
In this case, we resorted to directly giving \chatgpt\ the transition table, after which it performed the conversion correctly.
For all other questions that used Word's built-in equation editor, \chatgpt\ read the equation perfectly.

\chatgpt's approach for ``wrong'' answers was often very incremental, and never attempted to diagnose any core issue with its initial reasoning.
For example, Midterm 1's second question involved creating a small DFA for a given language.
\chatgpt's initial solution with 4 states was incorrect, but was fixed after a re-try.
The interesting part here is that \chatgpt's correct DFA still had 4 states, whereas the ``model'' solution only needed 3; here, the 4th state was unreachable and so was ``useless.''

In general, as expected, \chatgpt\ performed well at questions that are nearly identical to those that are posted online for which it could train on, and very poorly on questions that are not.
\chatgpt\ frequently ran into trouble even in the first case because small deviations in question phrasing can yield large changes in the answer as the underlying meaning could change.
For example, consider the question: ``True or False: the complement of the language $\{a^n b^n c^n : n \ge 0\}$ is context-free.'' 
The answer is true and is a standard exercise to prove it (as is similar to a true/false with explanation question in Midterm 2); however, if the ending was changed to ``deterministic context-free'', then the answer is false for a completely different reason.

Often the culprit for poorly performed questions is the notion of ``universality,'' in treating every case. 
For example, the Pumping Lemma question in both Midterms involved showing a given language is either not regular or not context-free; to adequately do so, all possible cases of decomposing a string needed to be considered. 
In neither case did \chatgpt\ handle all such cases.
Whenever a question is ``existential'' in that one needs to perform an algorithm on a given DFA, or give an example CFG, then \chatgpt\ performs much better.
\chatgpt\ performed better still when questions were phrased in a direct format without any ambiguity in the truth of the answer, e.g., ``Prove that $L$ is regular'' vs. ``True or False: $L$ is regular.''

\chatgpt's reasoning for proof-type problems is very rarely formal.
An example is on the Final Exam where a short answer question involved determining if there is an algorithm for solving a given problem, such as checking if a Turing Machine halts on all inputs within a fixed number of steps. 
Much of \chatgpt's reasoning here involved loosely arguing why there was no such algorithm because one needed to consider ``all inputs'', even though there was a bound on how many tape cells can be visited (thus not all inputs even need to be considered).
In a majority of cases, however, \chatgpt\ either corrected or partially addressed its answers here after a re-try.

\chatgpt\ sometimes treated parts of problems as ``black boxes'' even if they are referential to other parts. 
For example, the Final Exam had a Rice's theorem application question, where students needed to determine if that theorem applies. 
In this case, it does not because the input to the problem is a Turing Machine, and the language involved the TM accepting two different strings; a given one, and the machine's own description.
This second string forces Rice's theorem to not apply as it breaks one of the theorem's conditions. 
Like nearly all of our ToC students, \chatgpt\ treated the TM's own description as any other ordinary string not dependent on the input, and concluded the theorem could be applied.

There were a handful of cases where \chatgpt\ completely ``understood'' the prompt (i.e., the question was reproduced perfectly), and the exercise was standard, but nevertheless \chatgpt\ made a simple mistake.
For example, the first question of Midterm 2 involved creation of a CFG that involves breaking the problem down into usage of three key components: closure under union, such as $L_1 \cup L_2$ where both languages are context-free; nesting, such as $\{a^n b^m c^m d^n : n, m \ge 0\}$; and multiple parts, such as $\{a^n b^n c^m d^m : n, m \ge 0\}$.
\chatgpt\ treated the ``multiple part'' aspect of the question as a ``nesting'' one, although after a re-try it corrected the mistake.



\subsection{Experiment 2 Discussion}

Again for brevity, we will not discuss every question, but rather highlights.
\chatgpt\ had the highest average with DFA-related questions, and the lowest average with Regex-related questions. 
This could be due to the type of question in each category. 
The questions for DFA's ended up being largely definitions, with a few being practice problems, as it is the first topic within Section 1 of our ToC course, and its definition is more involved than the Regex definition.
Consequently, the problems for Regex were largely practice problems that required a larger amount of creativity and critical thinking, and not just surface-level application of the definitions of topics. 


There were rare cases where the explanation and reasoning were very close to being fully correct, but some small error lead to \chatgpt's final answer being wrong; this was implied in our discussion about Experiment 1. 
These cases occurred especially with the proof questions, as there are many more opportunities within proof steps to be wrong, particularly with induction, Pumping Lemma, and Undecidable reduction proofs. 

\chatgpt\ is often very malleable in its responses, which would make specific feedback prone to being interpreted as a hint, leading to \chatgpt\ changing its answer to fit what the feedback is saying. 
By being vague we attempt to ensure that the knowledge being tested reflects only what \chatgpt\ currently possesses.

\chatgpt\ sometimes altered its response with a worse score than its initial guess by adding useless information to the question's intent.
For example, one question asked to determine if a given CFG is ambiguous, which involved showing two leftmost derivations of the same string.
\chatgpt\ understood this, but produced two derivations of two different strings that were very similar to each other, but not the same.
In the re-try, \chatgpt\ gave the same output along with a derivation of another string that was completely unrelated to what the question was asking.

\subsection{Effects of Prompt Engineering}

To understand prompt engineering's influence, we determine its effects on the 30 NFA questions; Table~\ref{tab:results} shows that the average score on these questions was 3.23.
Using the same rubric as Experiment 2, we tested the same questions with the following initial prompt:

\begin{quote}
    I want you to respond like you are the preeminent computer science professor in the world. You will give intuitive and complete responses to my questions. You will be given a series of questions designed to test your ability to answer such questions. You will respond methodically and give your complete reasoning for every question. I will give you feedback with a score out of 4, and you will improve your answer using my feedback as a guide for how much your answer was incorrect.
\end{quote}

We found that the average score on these questions rose to 3.53.
On average, the true/false questions had an increase to 3.6 (vs. 3.4), the multiple-choice questions increased to 3.9 (vs. 3.3), and the proof questions increased to 3.1 (vs. 3.0). 
It is clear the prompt engineering had a much larger effect on the multiple-choice questions.
We believe this is the case because there are more options available vs. the true/false questions, but are not unconstrained like the proof questions, and the prompt gives appropriate context to how \chatgpt\ should phrase its solutions to the questions.

\section{Threats to Validity}\label{sec:threats}

The largest possible threat to the validity of the results is the accuracy of assigning scores based on the rubric, which was done by only the first and second authors.
We mitigated this threat by having the third author independently verify the scores' accuracy, as he has taught the course 8 times.

Another threat comes from a potential low quality of the tested questions. 
In order to mitigate this threat, we ensured that the questions came only from existing ToC resources, such as the course textbook, class notes, pre-class readings, and former exams; the latter three came from our institution. 

\section{Recommendations and Future Work}\label{sec:future_work}

First we give our recommendations for educators who want to either alter their ToC course to accommodate potential use of LLMs like \chatgpt\ by students, or to incorporate assessments utilizing LLMs. 
Since we have shown that \chatgpt\ can effectively pass a ToC course, we recommend keeping assessments that are closed-resources, such as in-person exams, as the majority of the course grade.
As our results show, an experienced ToC educator can still largely and safely use questions that have not been asked before that still test the same fundamental concepts and theorems.  
However, an essential part of a successful ToC course is for students to think critically about claims, and to verify them. 
In spirit of this, we recommend to include an assignment that involves asking an LLM to prove a claim from a previous assessment, and then have students point out every error within the LLM's output. 

As \chatgpt\ was rarely formal, we recommend requiring students to provide as-formal-as-reasonable proofs of their claims, especially for reductions such as undecidable proofs. 
Although not necessarily new advice, we highly recommend educators inspect students' homework submissions for key, specific details in their proofs.
\chatgpt\ often includes all of the ``boilerplate'' correctly in each proof type, but excludes important details, such as how a reduction from one language to another exactly works.
Further, we recommend educators to require students create non-trivial state diagrams as \chatgpt\ is unable to make them unless they are very small, and even these are directly written in ASCII.

Now we describe future work.
Certainly a future work item would be to use the same method to other LLMs; we predict similar behavior in that proof-type questions will be done poorly, whereas algorithmic-type questions will be done well. 
Of course, the topics we chose are only a small subset of topics within ToC; one could in principle test \chatgpt\ against topics such as computational complexity theory. 

In the advent of LLMs, it is ever more important to verify potential proofs of claims, especially mathematical ones.
For future work, one type of question would be to determine if a given proof is correct, and if not determine where the fundamental issue(s) is/are; this was one of our Final Exam questions. 
For this question, \chatgpt\ identified the correct proof step that caused the problem, but had incorrect reasoning as to why it was wrong. 

One potentially great project is to build a LLM specifically for ToC content, especially with explaining concepts, definitions, and breaking down assigned problems at an undergraduate level without giving away answers. 
From experience, students have most issues with starting proofs and often forget pre-requisite content, namely discrete math.
Even though textbooks do provide such needed content, such as Chapter 0 of Sipser \cite{sipser2013introduction}, our students historically do not read such resources.
Further, we would recommend adding functionality to such an LLM that can walk through some algorithmic process, such as NFA to DFA conversion.
Such a project would require a very large data set of accurate ToC content for training, but we are not aware of such a resource if it exists.

\section{Conclusion}\label{sec:conclusion}

In this paper we determined whether or not \chatgpt\ can pass a ToC course.
We accomplished this through two experiments: one based on our course's exams, and another based on a large set of questions we developed.
Overall, our experiments show that \chatgpt\ can ``pass'' the course with between a {\tt B-} and a {\tt B} average letter grade.
\chatgpt\ does suffer from being largely unable to prove claims on which it has not been trained, but can solve very standard exercises. 

\begin{acks}
The opinions in the work are solely of the authors, and do not necessarily reflect those of the U.S. Army, U.S. Army Research Labs, the U.S. Military Academy, or the Department of Defense.
\end{acks}

\bibliographystyle{ACM-Reference-Format}
\bibliography{sample-base}

\end{document}